\pdfoutput=1
\documentclass[sigconf]{acmart}
\acmDOI{}
\acmISBN{}
\acmYear{2019}
\acmConference{arXiv}{NY}{USA}
\makeatletter
\renewcommand\@formatdoi[1]{\ignorespaces}
\makeatother
\acmPrice{}
\setcopyright{rightsretained}
\usepackage{amsmath}
\usepackage{bm}
\usepackage{booktabs} %
\usepackage{color}
\usepackage{listings}
\usepackage{setspace}
\usepackage{enumitem}
\usepackage{balance}

\usepackage[utf8]{inputenc}
\usepackage{amsmath}
\usepackage{amssymb}
\usepackage{acronym}
\usepackage{xcolor}

\newcommand{\STATU}{\relax\ifmmode\triangle\else $\triangle$\fi}
\newcommand{\STATD}{\relax\ifmmode\triangledown\else $\triangledown$\fi}

\DeclareMathOperator*{\argmin}{argmin}

\begin{document}
\title{Self-Attentive Document Interaction Networks for Permutation Equivariant Ranking}

\author{
 Rama Kumar Pasumarthi, Xuanhui Wang, Michael Bendersky, Marc Najork}
\affiliation{\institution{Google}}
\email{{ramakumar,xuanhui,bemike,najork}@google.com}

\begin{abstract}

How to leverage cross-document interactions to improve ranking performance is an important topic in information retrieval (IR) research. However, this topic has not been well-studied in the learning-to-rank setting and most of the existing work still treats each document independently while scoring. The recent development of deep learning shows strength in modeling complex relationships across sequences and sets. It thus motivates us to study how to leverage cross-document interactions for learning-to-rank in the deep learning framework.
In this paper, we formally define the permutation-equivariance requirement for a scoring function that captures cross-document interactions. We then propose a self-attention based document interaction network and show that it satisfies the permutation-equivariant requirement, and can generate scores for document sets of varying sizes. 
Our proposed methods can automatically learn to capture document interactions without any auxiliary information, and can scale across large document sets. 
We conduct experiments on three ranking datasets: the benchmark Web30k, a Gmail search, and a Google Drive Quick Access dataset. Experimental results show that our proposed methods are both more effective and efficient than baselines.

\end{abstract}


\maketitle

\section{Introduction} \label{sec:intro}

Ranking is a central problem in many applications of information retrieval (IR) such as search, recommender systems, and question answering. The purpose of a ranking algorithm is to sort \emph{a set of items} into a ranked list such that the utility of the entire list is maximized.
For example in search, a set of documents are to be ranked to answer a user's query. The utility of the entire list highly depends on the top ranked documents.

Learning-to-rank employs machine learning techniques to solve ranking problems. 
The common formulation is to find a function that can produce scores for the list of documents of a query. The scores can then be used to sort the documents.
Many early attempts to learning-to-rank cast a ranking problem as regression or classification~\cite{burges2010ranknet,joachims2006training}. In such methods, the loss function being minimized incurs a cost for an incorrect prediction of relevance labels (``pointwise'' loss) or pairwise preferences (``pairwise'' loss). Such formulations are, however, misaligned with the ranking objective where the utility is often defined over the entire list of documents. Indeed, the so called ``listwise'' methods that optimize a loss function defined over the entire list have been shown to learn better ranking functions~\cite{quoc2007learning,wu2010adapting,cao2007learning}.

While much research has been devoted to the evolution of loss functions, the nature of the learned scoring function has largely remained the same: a univariate scoring function that computes a relevance score for a document in isolation.
How to capture cross-document interactions is the motivation behind several previous works~\cite{diaz2007regularizing,qin2009global,Dehghani:SIGIR2017,ai2018learning,bello2018seq2slate,ai2019ICTIR_Groupwise}.
Early methods such as the score regularization technique~\cite{diaz2007regularizing} and the conditional random field based models~\cite{qin2009global} use the similarity between documents to smooth or regulate ranking scores. These methods, however, assume the existence of document similarity information from another source such as document clusters. More recently, neural learning-to-rank algorithms~\cite{ai2018learning,bello2018seq2slate} and click models~\cite{Borisov+al:2016} capture document interactions using recurrent neural networks over document lists. These methods, however, belong to the \emph{re-ranking} setting because they assume that the input is an \emph{ordered list}, but not a \emph{set}.

Another work that investigates the effect of document interactions on ranking quality is RankProb~\cite{Dehghani:SIGIR2017}. It is a bivariate neural scoring function that takes a pair of documents as input and predicts the preference of one over the other. 
More recently, a more general framework was proposed in~\cite{ai2019ICTIR_Groupwise} to learn multivariate ``groupwise'' scoring functions (GSFs). Though being able to model document interactions, both methods are highly inefficient at inference time. %
These models suffer from a \textit{training-serving} discrepancy: the function learned during training is different from the scoring function used in serving. For example, average pooling over the bivariate function learned during training is used as the scoring function in RankProb during serving. For higher-order interaction models (such as GSFs), the pooling is over an intractable number of permutations, and hence, \emph{approximation} via sub-sampling is used, which worsens the training-serving discrepancy and makes the inference unstable.

In this paper, we identify a generic requirement for scoring functions with document interactions: permutation-equivariance. We analyze the existing approaches with respect to this requirement.
Based on the this requirement, we propose a class of neural network models and show that they not only satisfy this requirement precisely, but are also more efficient in modeling document interactions and do not have the training-serving discrepancy. Our proposed method is based on self-attention on the document level. It naturally captures the cross-document interactions via the self-attention mechanism. 
To the best of our knowledge, our work is the first to use it to model document interactions for learning-to-rank.

Our contributions can be summarized as follows: 
\begin{itemize}
  \item We propose the permutation-equivariance requirement for any document interaction model and analyze existing methods with respect to this requirement.
  \item We identify a generic class of permutation-equivariant functions, instantiate it using a self-attentive document interaction network, and incorporate it into learning-to-rank.
  \item We empirically demonstrate the effectiveness and efficiency of our proposed methods on both search and recommendation tasks using three data sets.
\end{itemize}

This paper is organized as follows. We begin with a review of the literature in Section~\ref{sec:related_work}, and formalize the problem we wish to study in Section~\ref{sec:ltr}. In Section~\ref{sec:methods}, we present a detailed description of our proposed methods. We examine the effectiveness of our methods empirically and summarize our findings in Section~\ref{sec:experiments}. Finally, we conclude this work and discuss future directions in Section~\ref{sec:conclusion}.

\section{Related Work} \label{sec:related_work}

In learning-to-rank literature, a common approach is called ``score and sort''. For capturing the loss between the list of scores for documents and relevance labels, \textit{pointwise}~\cite{Fuhr:1989:OPR:65943.65944, Chu:2005:PLG:1102351.1102369}, \textit{pairwise}~\cite{burges2005learning, burges2010ranknet} and \textit{listwise}~\cite{xia2008listwise, xia2008listmle, bruch2019revisiting} losses have been extensively studied. Scoring functions have been parameterized by boosted decision trees~\cite{ke2017lightgbm}, SVMs~\cite{Joachims:2002}, and neural networks~\cite{burges2010ranknet, DeepRank:2017, pasumarthi2019tf}.

In the context of scoring query-document pairs, the recent neural ranking models have been broadly classified~\cite{guo2019deep} into two categories: \textit{representation} focused and \textit{interaction} focused. The methods that are representation-focused~\cite{Huang+al:2013, DeepRank:2017, pang2016text} look at learning optimal vector space representations for queries and documents, and then combine them using dot product or cosine similarity. The interaction-focused methods learn a joint representation based on interaction networks between queries and documents. These approaches, along with hybrid variants between representation and interaction focused~\cite{mitra2017learning}, are \textit{univariate} approaches, i.e., they deal with scoring a query-document pair, and do not capture cross-document interactions. Please note that attention mechanism~\cite{romeo2016neural} has also been explored in this line of work, but it is mainly used in the word or paragraph level, not the document level.

Recent work in modeling document interactions in learning-to-rank have focused on the re-ranking scenario~\cite{ai2018learning, bello2018seq2slate, pei2019personalized}, where the input is an ordered list of documents, not a set. These are not applicable to full set ranking, which is the focus of our work. Regularizing scores~\cite{diaz2007regularizing}, and a CRF approach~\cite{qin2009global} using document cluster information to augment the training loss have been explored, which are complementary to our proposed approach.

\section{Problem Formulation} \label{sec:ltr}

\newcommand{\cD}{\mathcal{D}}
\newcommand{\bd}{\bm{d}}
\newcommand{\bx}{\bm{x}}
\newcommand{\by}{\bm{y}}
\newcommand{\hby}{\hat{\bm{y}}}

In this section, we formulate our problem in the learning-to-rank setting. 

\subsection{Learning-to-Rank}

Learning-to-rank solves ranking problems using machine learning techniques. In such a setting, the training data consists of a set of queries with each having a list of documents that we would like to rank. Formally, let $\cD=\{(q, \bd, \by)\}$ be a training data set where $q$ is a query, $\bd$ is the list of documents for $q$, and $\by$ is the relevance labels for $\bd$. We use $d_i$ and $y_i$ to refer to the $i$-th elements in $\bd$ and $\by$ respectively. A scoring function $s$ takes both $q$ and $\bd$ as input and computes a vector of scores $\hby$:
\begin{equation} \label{eq:scoring_function}
    \hby = s(q, \bd).
\end{equation}
A loss function $\ell$ for query $q$ can be defined between the predicted scores and the labels:
$$\ell(q, \bd, \by) = \ell(\by, \hby)$$
The goal of a learning-to-rank task is to find a scoring function $s^*$ that minimizes the empirical loss over the training data:
\begin{equation} \label{equ:empirical_loss}
s^* = \argmin_{s\in\mathcal{H}} \frac{1}{|\cD|}\sum_{(q, \bd, \by) \in \cD} \ell(q, \bd, \by).
\end{equation}
Typical examples of the hypothesis space $\mathcal{H}$ for a scoring function $s$ are support vector machines~\cite{joachims2006training, Joachims:WSDM17}, boosted weak learners~\cite{Jun+Hang:2007}, gradient-boosted trees~\cite{friedman2001greedy, burges2010ranknet}, and neural networks~\cite{burges2005learning}.

\subsection{Ranking Loss Functions}
Given a formulation of the scoring function, there are various definitions of ranking loss functions~\cite{liu2009learning}. In this paper, we focus on the following two listwise loss functions as they have been shown to be closely related to the ranking metrics~\cite{bruch:ictir2019, Qin:2010:GAF:1842549.1842572, bruch2019revisiting}. The first one is the Softmax Cross-Entropy loss (denoted as Softmax) and has been shown to be a proper bound of ranking metrics over binary relevance labels like MRR~\cite{bruch:ictir2019}:
\begin{align}\label{eq:softmax}
    \ell(\by, \hby) = \sum_{i} \frac{y_{i}}{\sum_{i'} y_{i'}} \log\Big(\frac{\exp(\hat{y}_{i})}{\sum_{i'} \exp(\hat{y}_{i'})}\Big).
\end{align}
where the subscript $i$ and $i'$ means the $i$-th or $i'$-th element in a vector.

The second one is the Approximate NDCG loss (denoted as ApproxNDCG)~\cite{qin2010general, bruch2019revisiting}. It is more suitable for graded relevance labels, and is derived from the NDCG metric, but uses scores to approximate the ranks to make the objective smooth:
\begin{align}\label{eq:approxndcg}
    \ell(\by, \hby) = \frac{1}{DCG^*(\by)} \sum_{i} \frac{2^{y_{i}} - 1}{\log_2(1 + \hat{r}_{i})},
\end{align}
where $DCG^*(\by)$ is the normalization term of NDCG and $\hat{r}_{i}$ is the approximate rank defined as 
\begin{align*}
    \hat{r}_{i} = 1 + \sum_{i': i' \neq i} \frac{\exp(-\eta \hat{y}_{i'})}{\exp(-\eta \hat{y}_{i}) + \exp(-\eta \hat{y}_{i'})},
\end{align*}
where $\eta$ is the parameter that controls the closeness of the approximation. When $\hat{r}_i$ is replaced by the rank $r_i$ sorted by scores $\hat{y}_i$, Equation~\ref{eq:approxndcg} becomes the NDCG metric. A larger $\eta$ makes $\hat{r}_i$ closer to $r_i$, but it also makes ApproxNDCG less smooth and thus harder for optimization. We tune $\eta$ in our experiments and set $\eta=0.1$ since it gives the optimal results. 

\subsection{Permutation-Equivariance Requirement}

Our focus in this paper is on scoring functions. We postulate that it is preferable that the scoring function is \textit{permutation equivariant}, so that the resulting ranking will be independent of the original ordering induced over the items by the candidate generation process (e.g., a base ranker, or a retrieval mechanism). This ensures that the learned ranker will not be biased by any errors in the candidate generation process.
We first give the general mathematical definition of \emph{permutation-equivariant} functions.

\begin{definition}[Permutation-Equivariant Functions] \label{def:permutation-equvariant}
Let $\bx$ be a vector of elements $[x_1, ..., x_n]$, where $x_i \in \mathcal{X}$, and $\pi$ is a permutation of indices $[1,..., n]$. A function $f: \mathcal{X}^n \rightarrow \mathcal{Y}^n$ is \emph{permutation-equivariant} iff
$$f(\pi(\bx)) = \pi(f(\bx)).$$
That is, a permutation applied to the input vector will result in the same permutation applied to the output of the function.
\end{definition}

For a scoring function $s(q, \bd)$, the input domain $\mathcal{X}$ is defined by the representation of $q$ and $d_i$ (e.g., $\mathbb{R}^{k_q + k_d}$ where $k_q$ and $k_d$ are the dimension of their vector representation) and the output domain is $\mathcal{Y}=\mathbb{R}$. It is permutation-equivariant iff 
$$s(q, \pi(\bd)) = \pi(s(q, \bd)).$$
We analyze some existing work in term of this requirement. 

The vast majority of learning-to-rank algorithms assume a \emph{univariate} scoring function that computes a score for each document independently of others. With slight abuse of notation, we also use $s$ to represent the scoring function on each individual document:
\begin{equation} \label{eq:univariate}
    \hat{y}_i = s(q, d_i)
\end{equation}
where $d_i$ is an individual document in the list $\bd$ and $\hat{y}_i$ is the $i^{th}$ value of the score vector $\hby$. A univariate scoring function is permutation-equivariant because
\begin{align*}
\pi(s(q, \bd))& = \pi([s(q, d_1), s(q, d_2), ...]) \\
& = [s(q, d_{\pi(1)}), s(q, d_{\pi(2)}), ...] \\
& = s(q, \pi(\bd)).  
\end{align*}

The Groupwise Scoring Functions (GSFs)~\cite{ai2019ICTIR_Groupwise} boil down to univariate scoring functions when the group size is 1. A larger group size is needed to model cross-document interactions. For example, for groups of size 2, the scoring of the $i$-th document is:
\begin{equation}\label{eq:pair-pooling}
\hat{y}_i = \frac{1}{2(n-1)}\sum_{d_j} g(q, d_i, d_j) + g(q, d_j, d_i),
\end{equation}
where $g$ is the sub-scoring function in GSF and is implemented using feed forward networks. Higher-order interactions are explicitly captured when the group size is larger, but it becomes impractical to implement a GSF precisely due to the combinatorial number of groups. Monte Carlo sampling methods are used to approximate and this can make GSFs unstable. In this sense, GSFs are \emph{approximately} permutation-equivariant.

The RankProb approach in~\cite{Dehghani:SIGIR2017} trains a bivariate interaction scoring function $g(q, d_i, d_j)$ by concatenating the features as the input for a feed forward network. The loss function is a logistic regression on the pairwise preference of the two documents. For inference, it uses the average pooling in Equation~\ref{eq:pair-pooling}. This model is similar to the GSFs with group size 2. It has a training-serving discrepancy. The average pooling makes the scoring function permutation-equivariant but directly using it has a $O(n^2)$ time complexity, which is not scalable.

\begin{figure}
\centering
\includegraphics[width=0.9\linewidth,height=\textheight,keepaspectratio]{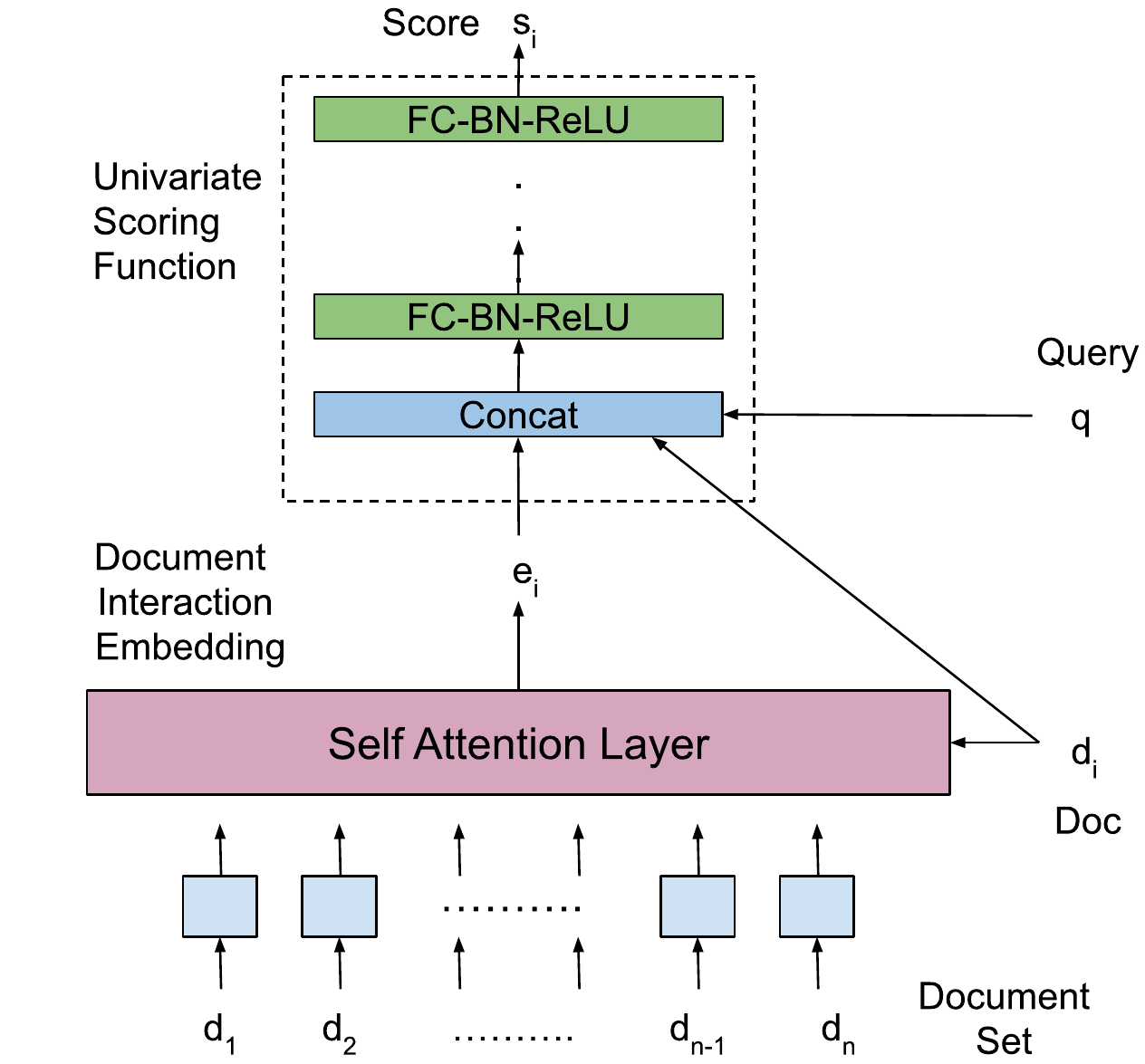}
\vspace{-5pt}
\caption{\small Self-Attentive Document Interaction Network.}
\label{fig:attn_din_arch}
\vspace{-5pt}
\end{figure}
\newcommand{\R}{\mathbb{R}}
\section{Proposed Methods} \label{sec:methods}

In this section, we first present a general class of permutation-equivariant functions and outline how we build a permutation equivariant scoring function using deep neural networks for our proposed approach.

\subsection{A Class of Permutation Equivariant Functions}

Our permutation-equivariant functions are based on permutation-invariant functions. We start with the formal definition of permutation-invariant functions.

\begin{definition}[Permutation-Invariant Functions] \label{def:permutation-invariant}
Let $\bx$ be a vector of elements $[x_1, ..., x_n]$, where $x_i \in \mathcal{X}$ and $\pi$ be a permutation of indices of $[1,..., n]$. A function $f: \mathcal{X}^n \rightarrow \mathcal{Y}$ is \emph{permutation-invariant} iff
$$f(\pi(\bx)) = f(\bx).$$
That is, any permutation of the input $\bx$ has the same output.
\end{definition}

The work in~\cite{zaheer2017deep} provides a general characterization of permutation-invariant functions as follows:

\begin{theorem} \label{thm:permutation-invariance}
A function $f(\bx)$ is \emph{permutation-invariant} iff it can be decomposed in the form $\rho \left( \sum_{x_i \in \bx} \phi(x_i) \right)$, for a suitable choice of of $\rho$ and $\phi$.
\end{theorem}

Though simple, Theorem~\ref{thm:permutation-invariance} is less constructive. Ilse et. al.~\cite{ilse2018attention} proposed a mechanism to extend the form in Theorem~\ref{thm:permutation-invariance} (called \textit{pooling} function) to a weighting pooling, based on the attention mechanism. We shall refer this as \textit{attention pooling} function. Given a generic context $c$, a pooling function can be extended to attention pooling as follows:
\begin{equation} \label{eq:attention-pooling}
f(\bx; c) = \rho \left( \sum_{x_i \in \bx} \alpha(x_i, c)~\phi(x_i) \right)
\end{equation}
Here, $\alpha(.)$ is the popular attention mechanism, which is used to capture the similarity between the context and the item.

The class of permutation-equivariant functions in this paper is based on self-attention~\cite{lin2017structured}. The key idea is to instantiate the context $c$ by an item $x_i$ in $\bx$. Based on Equation~\ref{eq:attention-pooling}, we form a function $F: \mathcal{X}^n \rightarrow \mathcal{Y}^n$ that can be verified to be permutation-equivariant as follows:
\begin{align} \label{eq:self-attention}
F\left(\bx \right) &= \{f(\bx; c=x_k)\}_{k=1}^n \\ \nonumber
& = \left\{\rho \left( \sum_{x_i \in \bx} \alpha(x_i, x_k)~\phi(x_i)\right)\right\}_{k=1}^n 
\end{align}

\subsection{Self-Attentive Document Interaction Networks} \label{sec:din}
We instantiate the permutation-equivariant functions using the sclaed-dot product attention, proposed in the work on Transformer~\cite{vaswani2017attention}. 

\subsubsection{Self-Attention Layers}
The attention layer in Transformer is defined based on three matrices: $Q \in \R^{n_q \times k}, K \in \R^{n_k \times k}, V \in \R^{n_k \times n_v}$, where $k$ is the dimension of keys in $K$ matrix, as follows:
\begin{align} 
  Attention(Q, K, V) := softmax(\frac{QK^T}{\sqrt{k}}) V 
\end{align}
The output of the attention is a matrix in $\R^{n_q \times n_v}$. The self-attention is a special case of the attention where we use $Q=K=V$. In our setting, we implement each row of $V$ as the concatenation of the vector representation of $q$ and each $d_i$. The self-attention is permutation-equivariant by using each row of $V$ as  $\phi(x_i)$ and setting the matrix form of $\alpha$ as $softmax(\frac{VV^T}{\sqrt{k}})$ in Equation~\ref{eq:self-attention}.
Similar to the work on Transformer~\cite{vaswani2017attention}, we use layer normalization~\cite{ba2016layer} and residual connections~\cite{he2016deep} over the output of the self-attention and these operations form the function of $\rho$ in Equation~\ref{eq:self-attention}.

Furthermore, we use the multi-headed attention mechanism, which allows the model to learn multiple attention weights per layer:
\begin{align} 
    MultiHead(Q, K, V) &:= concat(head_1, ..., head_n) W^O   \\
    head_i &:= Attention(Q W_i^Q, K W_i^K, V W_i^V) \nonumber
\end{align}
where matrices $W_i$'s are the weight matrices for each head. Heads are concatenated along rows and projected by $W^O$. Again we can have a self-attention layer by setting $Q=K=V$ and show this is permutation-equivariant.

We note that such an self-attention mostly take the pairwise document interactions. Since permutation-equivariance is preserved for function composition $(G\circ F)(\bx) = G(F(\bx))$, we can stack multiple self-attention layers. Multiple layers can enhance and potentially capture higher-order interactions better.

\subsubsection{Scoring Layers}
However, our goal is to derive a permutation-equivariant scoring function whose output is $\mathcal{Y}^n = \R^n$. We propose to use a univariate scoring function $s$ on top of the output of self-attention layers. Let $F(q, \bd)$ be the output of self-attention layers and $e(d_i; \bd) = F(q, \bd)_{i}$ be the $i$-th row of the output, corresponding to document $d_i$.
We propose a ``wide and deep'' scoring function to combine self-attention based features with query and document features:
$$\hat{y}_i = s(q, d_i, \bd) = s(q, d_i, e(d_i; \bd))$$
We refer to this a as ``wide and deep'' architecture, where the output of ``deep'' layers, a stack of self-attention layers, is combined in a ``wide" univariate scoring function with query and document features to generate a score per document. 

We show that this ``wide and deep'' scoring function (denoted as $s_{DIN}$) is still permutation-equivariant, while capturing cross-document interactions.
\begin{align*}
\pi(s_{DIN}(q, \bd))& = \pi([s(q, d_1, e(d_1; \bd)), s(q, d_2, e(d_2; \bd))), ...]) \\
& = [s(q, d_{\pi(1)}, e(d_{\pi(1)}; \bd))), s(q, d_{\pi(2)}, e(d_{\pi(2)}; \bd))), ...] \\
& = [s(q, d_{\pi(1)}, F(q, \bd)_{\pi(1)}), s(q, d_{\pi(2)}, F(q, \bd)_{\pi(2)}), ...] \\
& = s_{DIN}(q, \pi(\bd)).  
\end{align*}

We call our method Self-Attentive Document Interaction Networks (denoted as attn-DIN) and the structure of the score for a given document is shown in Figure~\ref{fig:attn_din_arch}. The self-attention layer can be stacked sequentially, without losing the permutation equivariance property. In ``wide and deep'' fashion, the output of this layer, document interaction embeddings, is combined with query and document features and fed as an input to a univariate scoring function. 
The specific univariate scoring function captures interactions between features using a stack of feedforward layers. Specifically, for each feedforward layer, the input is passed through a dropout regularization~\cite{srivastava2014dropout} (to prevent overfitting), and the output is passed through a batch normalization layer~\cite{ioffe2015batch}, followed by a non-linear ReLU~\cite{nair2010rectified} activation, where $ReLU(x) = \max(x, 0)$. We refer to this combination as \textit{FC-BN-ReLU} in Figure~\ref{fig:attn_din_arch}. The final output is projected to a single score for a document.

\begin{table*}[]
\centering
\caption{\small Comparison of NDCG between GSF models and attn-DIN for cross-document interactions on Web30k data. $^*$ indicates the best GSF model. $\STATU$/$\STATD$ indicate statistically significant increase/decrease compared to best GSF model (p-value<0.05).}
\begin{tabular}{@{}llll@{}}
\toprule
Method                      & NDCG@1           & NDCG@5               & NDCG@10           \\ \midrule
GSF(m=64) with Softmax (best reported~\cite{ai2019ICTIR_Groupwise})       & 44.21 & 44.46  & 46.77  \\  
GSF(m=1) with ApproxNDCG (best reported~\cite{bruch2019revisiting})     & 46.64  & 45.38 & 47.31  \\  
GSF(m=1) with ApproxNDCG (finetuned)$^*$        & 46.81  & 45.59 & 47.39  \\
attn-DIN with ApproxNDCG (proposed approach)    & \bf{48.16}$^{\STATU}$  & \bf{46.62}$^{\STATU}$ & \bf{48.21}$^{\STATU}$   \\ \midrule
LambdaMART (RankLib)                     & \it{45.35}  & \it{44.59}  & \it{46.46} \\             
LambdaMART (lightGBM)                      & \it{50.33}  & \it{49.20}  & \it{51.05}  \\  \bottomrule
\end{tabular}
\label{tab:web30k_metrics}
\end{table*}

\section{Experiments}\label{sec:experiments}

In this section,
we first outline several learning-to-rank datasets and baseline methods used in our experiments. We then report the comparisons on both model effectiveness and inference efficiency.
 \vspace{-5pt}
\subsection{Datasets} \label{sec:datasets}

\subsubsection{MSLR Web30k}
The Microsoft Learning to Rank (MSLR) Web30k~\cite{quoc2007learning} public dataset comprises of 30,000 queries. We use Fold1, which has 3 partitions: train, validation, and test. For each query-document pair, it has 136 dense features. Each query has a variable number of documents, and we use at most 200 documents per query for training baseline and proposed methods. During evaluation, we use the test partition and consider all the documents present for a query. We discard queries with no relevant documents both during training and evaluation. 

\subsubsection{Quick Access}
In Google Drive, Quick Access~\cite{tata2017quick} is a zero-state recommendation system, that surfaces relevant documents that users might click on when they visit the Drive home. The features are all dense, as described in Tata et. al.~\cite{tata2017quick}, and user clicks are used as relevance labels. We collect around 30 million recommended documents and their click information. Each session has up to 100 documents, along with user features as contextual information for training and evaluation. We use a 90\%-10\% train-test split on this dataset.

\subsubsection{Gmail Search}
In Gmail, we look at search over e-mails, where a user types in a query, looks for a relevant e-mail, and clicks on one of the six results returned by the system. The list of six e-mails are considered as the candidate set, and the clicks are used as the relevance labels. To preserve privacy, we remove personal information, and apply $k$-anonymization. Around 200 million queries are collected, with a 90\%-10\% train-test split. The features comprise of both dense and sparse features, The sparse features comprise of character and word level n-grams with $k$-anonymization applied.

\vspace{-5pt}
\vspace{-5pt}
\subsection{Baselines} \label{sec:baselines}
On the public Web30k dataset, we compare with LambdaMART's implementation in RankLib~\cite{croft2013lemur} and lightGBM~\cite{ke2017lightgbm}, and with multivariate Groupwise Scoring Functions~\cite{ai2019ICTIR_Groupwise}. Since the labels consist of graded relevance, for evaluation measures, we use Normalized Discounted Cumulative Gain (NDCG)~\cite{jarvelin2002cumulated} for top 1, 5, and 10 documents ordered by the scores. 

On the private datasets of Quick Access and Gmail, we compare only with Groupwise Scoring Functions. Given the massive scale of the datasets, and the heterogeneous nature of features (dense and sparse), the open source implementations of LambdaMART do not scale on these datasets. Furthermore, prior work demonstrated that GSFs are superior to LambdaMART when sparse features are present~\cite{ai2019ICTIR_Groupwise}. We evaluate using Mean Reciprocal Rank~\cite{craswell2009mean} and Average Relevance Position~\cite{zhu2004recall}, as the labels are binary clicks, for which these two measures are most suitable.

\vspace{-5pt}
\subsection{Hyperparameters} \label{sec:hyperparameters}
On Web30k dataset, to encode document interactions, we use one self-attention layer with 100 neurons, and with a single attention head.
The univariate scoring function to combine the output of self-attention with query and document features comprises of an input batch normalization layer, followed by 7 feedforward fully-connected layers (\textit{FC-BN-ReLU} layers, as shown in Figure~\ref{fig:attn_din_arch}) of sizes $1024,512,256,128,64,32,16$. The model is trained using a training batch size of 128, and Adagrad~\cite{duchi2011adaptive} optimizer, with a learning rate of 0.005 to minimize the ApproxNDCG ranking loss. 
We use a similar setup for Gmail and Quick Access, with Softmax loss minimized using Adagrad Optimizer, trained for 10 million and 5 million steps respectively. For Gmail, we use 5 layers of self-attention with 100 neurons each, with 4 heads for encoding document interactions. For Quick Access, we use 3 layers of self-attention with 100 neurons each, with 5 heads for encoding document interactions.

\begin{figure*}
\centering
\begin{minipage}{0.32\textwidth}
\includegraphics[width=\linewidth,height=\textheight,keepaspectratio]{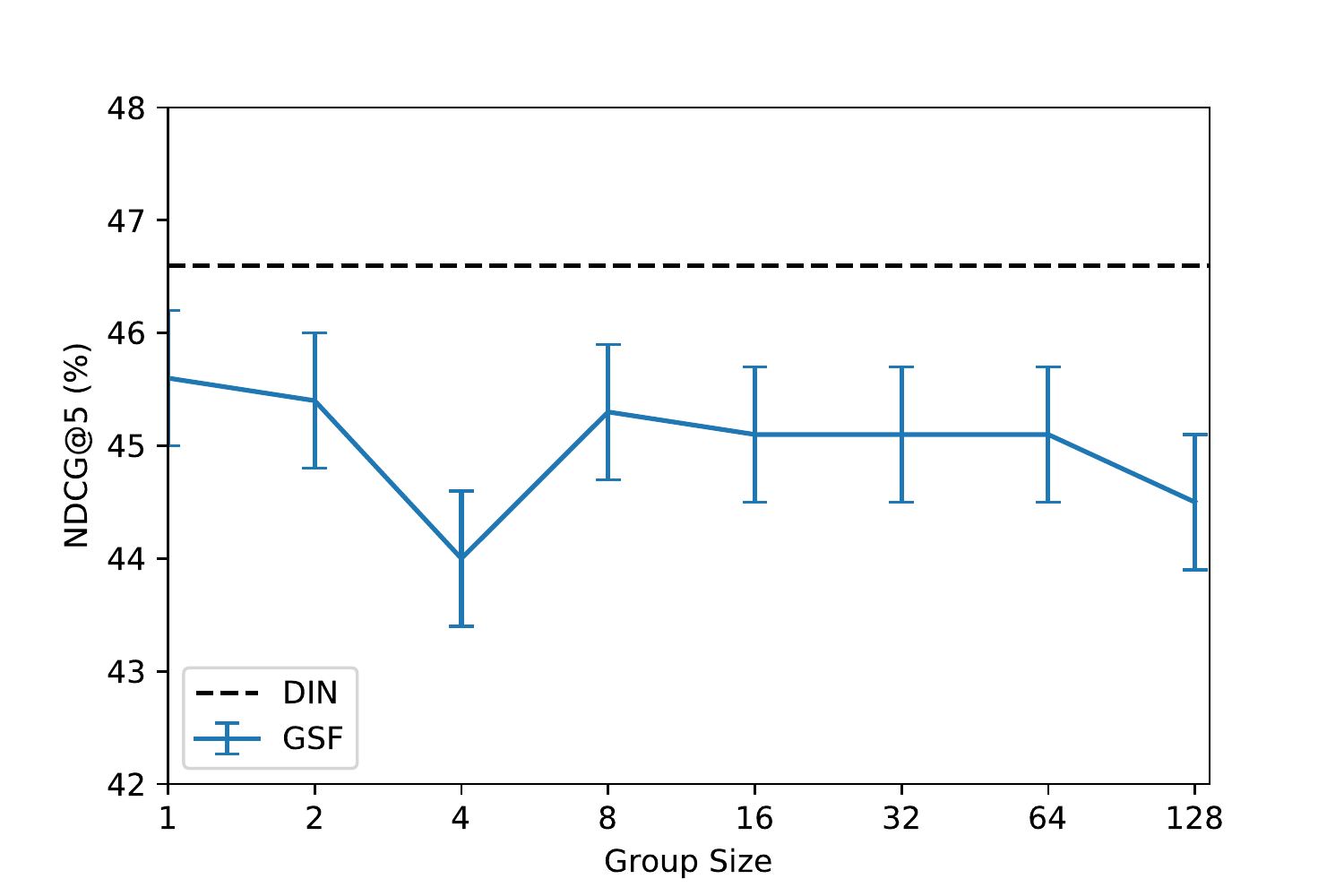}
\end{minipage}
\begin{minipage}{0.32\textwidth}
\centering
\includegraphics[width=\linewidth,height=\textheight,keepaspectratio]{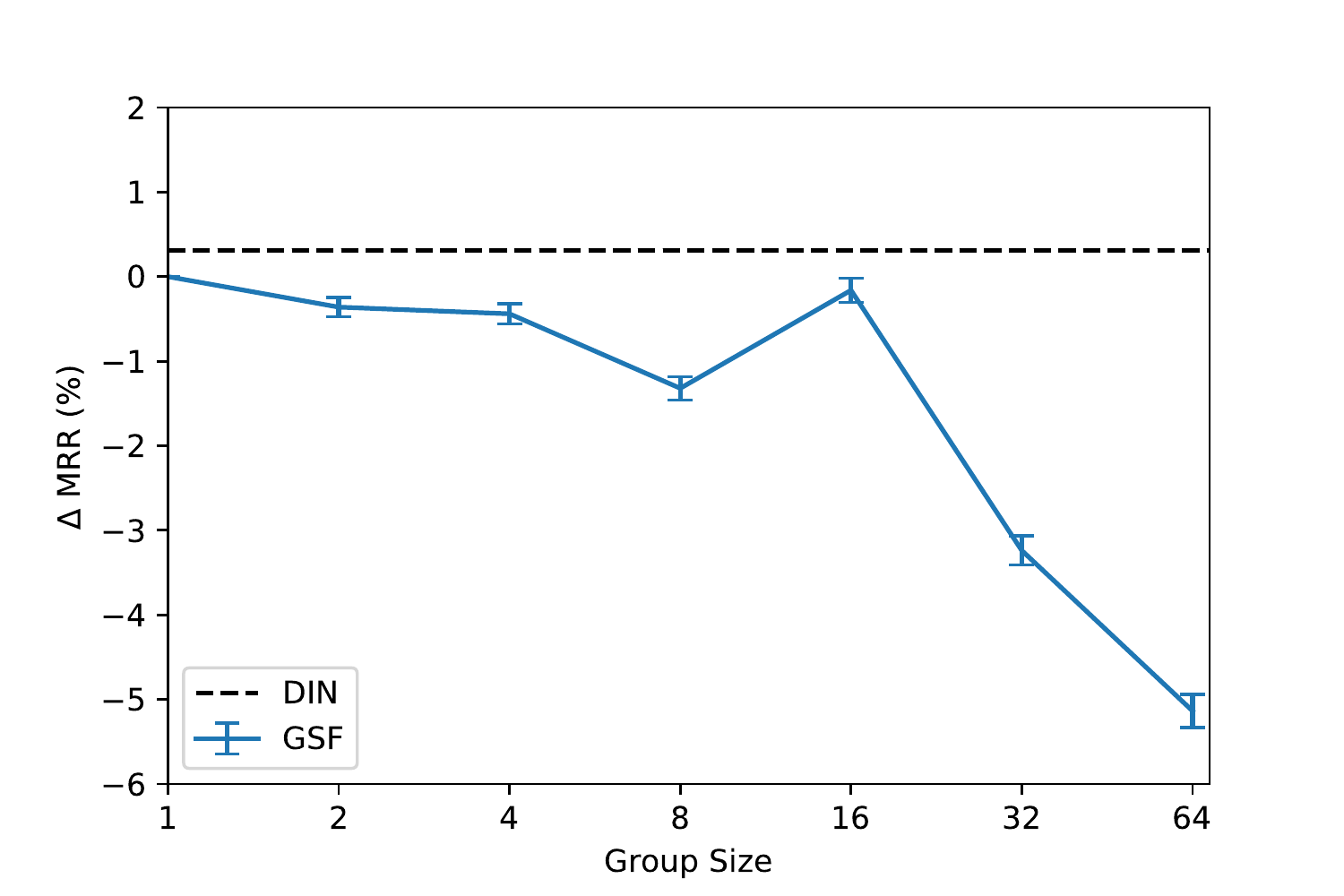}
\end{minipage}
\begin{minipage}{0.32\textwidth}
\centering
\includegraphics[width=\linewidth,height=\textheight,keepaspectratio]{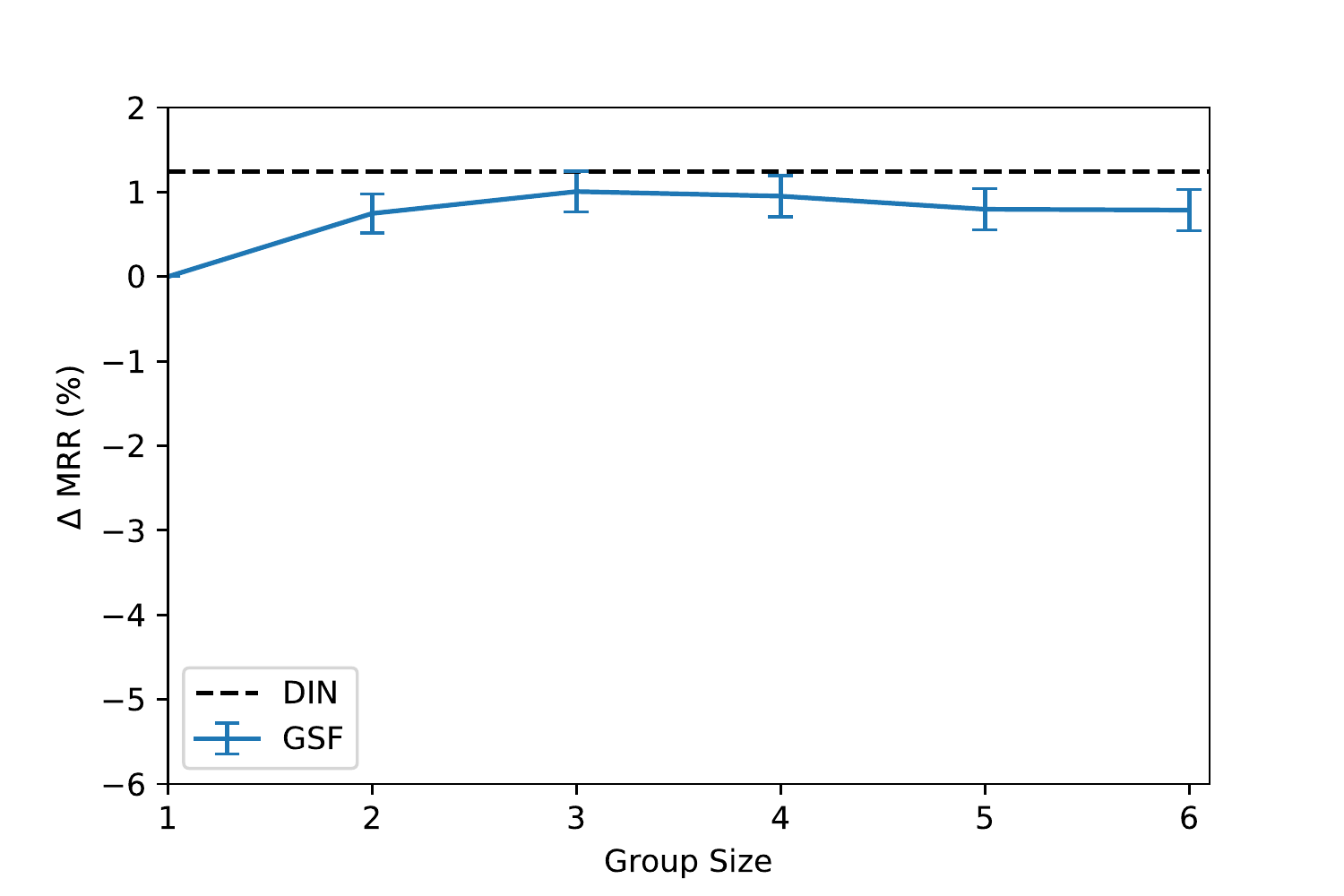}
\end{minipage}
\vspace{-5pt}
\caption{\small Comparison of ranking metrics between GSF models and attn-DIN on (left to right): Web30k, Quick Access and Gmail datasets.}
\label{fig:metrics_gsf_din}
\vspace{-5pt}
\vspace{-5pt}
\end{figure*}

\subsection{Model Effectiveness}

\begin{table}\small
\caption{\small Model performance on Quick Access and Gmail data. Note that $\Delta$MRR and $\Delta$ARP denote \% relative improvement. 
The best performance per column is in bold. $^*$ indicates the best GSF model. $\STATU$/$\STATD$ indicate statistically significant increase/decrease compared to the best GSF model (p-value<0.05).}
\vspace{-10pt}
\label{tab:qa_gmail_metrics}
\centering
\begin{tabular}{@{}lll@{}}
\toprule
(a) Quick Access                      & $\Delta$MRR      &  $\Delta$ARP      \\ \toprule
GSF(m=1$)^*$ (univariate scoring)                          & --               & --                \\
GSF(m=4)                                   &  -0.440 $\pm$ 0.177$^{\STATD}$                       & -0.659 $\pm$ 0.141$^{\STATD}$     \\
attn-DIN (proposed approach)               & \bf{+0.312} $\pm$ 0.113$^{\STATU}$ & \bf{+0.413} $\pm$ 0.124$^{\STATU}$ \\ \midrule \midrule
(b) Gmail Search                      & $\Delta$MRR      &  $\Delta$ARP      \\ \midrule
GSF(m=1) (univariate scoring)         &   --         & --                     \\    
GSF(m=3$)^*$                                & +1.006 $\pm$ 0.247 & +1.308 $\pm$ 0.246 \\
attn-DIN (proposed approach)               &  \bf{+1.245} $\pm$ 0.228$^{\STATU}$ & \bf{+1.430} $\pm$ 0.247 \\ \bottomrule
\end{tabular}
\vspace{-5pt}
\end{table}

\begin{figure}
\centering
\includegraphics[width=\linewidth,height=\textheight,keepaspectratio]{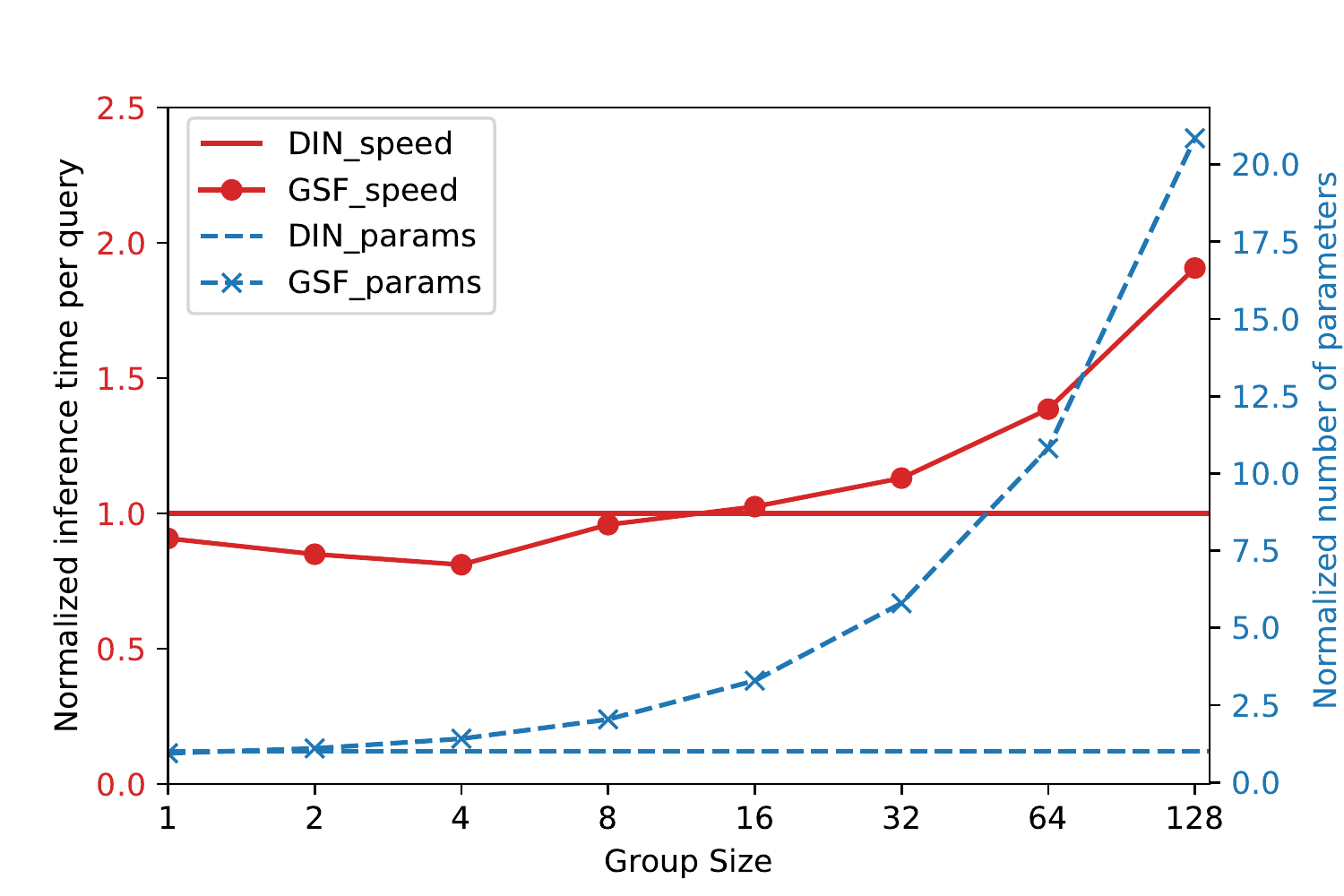}
\caption{\small Comparison of normalized inference time per query and number of parameters between GSF models and attn-DIN.}
\label{fig:inference_num_params}
\vspace{-15pt}
\end{figure}

In Table~\ref{tab:web30k_metrics}, on Web30k data, we compare the proposed attn-DIN approach with LambdaMART and GSFs. For LambdaMART, we consider the lightGBM implementation and the older RankLib implementation, and list the best reported metrics on test data. For the GSFs, we list the best reported metrics, and also an improved model based on our finetuning experiments. 
In Figure~\ref{fig:metrics_gsf_din}, we compare attn-DIN model with multivariate GSF models for varying group sizes. Since the list size is large for Web30k dataset (around 200), we increase the group size on an exponential scale from 1 to 128. We also show 95\% bootstrapped confidence intervals for each of the models.

We observe that the proposed approach significantly outperforms both the best reported and finetuned GSFs, giving around 1 point improvement for NDCG@5 (measured from 0-100), which is statistically significant ($+1.03 \pm 0.35$), measured using paired $t$-test with p-value threshold of $0.05$. These gains are not just from using a deeper network or from more neural network parameters, as shown in Figure~\ref{fig:inference_num_params}. The increase in number of parameters over the univariate scoring is the smallest for attn-DIN model, compared to any of the GSF models, while the improvement in ranking measure is significant. Our attn-DIN model tries to capture similarity using dot product attention mechanism and pooling to combine feature values, while GSFs explicitly model cross-document interactions using feedforward networks. As the group size increases, the number of parameters needed to capture cross-document interactions also increase. This also leads to increase in inference time, as discussed in Section~\ref{sec:inference_time}. 

The proposed approach outperforms RankLib's LambdaMART, but not the lightGBM implementation. We believe this is due to the fact that Gradient Boosted Decision Trees are very powerful class of machine learning models when the feature set consists of purely dense features, and smaller training datasets.

In most real world scenarios, input features tend to have both dense and sparse features. Query, document titles and metadata tend to naturally have textual description, which play a key role during user's relevance judgment, and hence are powerful signals for training ranking models. We look at two real world datasets, on Gmail Search and Quick Access, with a large amount of data and a variety of features, as described in Section~\ref{sec:hyperparameters}. In Table~\ref{tab:qa_gmail_metrics}, we report relative improvements in MRR, due to  private nature of these datasets.
For statistical significance, we use paired \textit{t}-test over relative improvements in MRR, with p-value threshold of $0.05$.

On the Quick Access dataset (Table~\ref{tab:qa_gmail_metrics}(a)), we analyze the relative improvements in MRR, and observe that the proposed approach does significantly better than the univariate model, which is in fact, the best performing GSF model. While the GSF models fail to produce any improvements from cross-document interactions on this dataset, our proposed approach effectively captures them.

On the Gmail dataset (Table~\ref{tab:qa_gmail_metrics}(b)), the proposed approach is significantly better than the univariate model, and is superior to the best GSF model ($m=3$). We conducted a paired \textit{t}-test between attn-DIN and the best GSF model, and we observe a relative improvement in MRR, $+0.237 \pm 0.206$, which is a statistically significant improvement. Note that in Gmail, we consider smaller document candidate sets (6 document per query), whereas in Web30k and and Quick Access, we use much larger candidate sets (200 documents per query and 100 documents per user request, respectively). For larger group sizes (> 8), the performance of GSF models deteriorates, whereas the proposed approach is able to capture cross-document interactions effectively.

\vspace{-5pt}
\subsection{Model Efficiency} \label{sec:inference_time}
In Figure~\ref{fig:inference_num_params}, we compare the inference time and number of parameters for various GSF models, normalized with the value for the proposed approach. Over univariate scoring functions, the proposed approach has an increase in inference time similar to GSF model of group size 8, despite capturing interactions across the entire document set of sizes 200 for Web30k. For the GSF models, the inference time increases with the increase of group size. GSFs use an approximation during inference. For group size 2, it uses a rolling window of 2 over a shuffled list to reduce the time complexity to $O(n)$~\cite{ai2019ICTIR_Groupwise}. However, it is not guaranteed to be permutation-equivariant, and may be unstable during inference. The exact inference is using Equation~\ref{eq:pair-pooling}, the same as RankProb~\cite{Dehghani:SIGIR2017}, which has $O(n^2)$ time complexity. In our experiments, it takes around $2,240$ ms for inference per query, and is drastically slower than the attn-DIN approach, which takes around $13$ ms for inference per query.  

For the Web30k dataset, from Figures~\ref{fig:metrics_gsf_din} and ~\ref{fig:inference_num_params}, we can observe that the proposed approaches are significantly better than univariate approaches, and are faster during inference than GSF models at large group sizes while capturing cross-document interactions across the list. From Figure~\ref{fig:inference_num_params}, we can also observe that the proposed model has fewer parameters than multivariate GSF models; hence the gain in ranking metrics is not from using larger number of parameters, but from effectively capturing similarity via cross-document attention pooling of the document features.

\vspace{-5pt}
\section{Conclusion} \label{sec:conclusion}
In this paper, we study how to leverage document interactions in the learning-to-rank setting. We proposed the permutation-equivariance requirement for a scoring function that takes document interactions into consideration. We show that self-attention mechanism can be used to implement such a permutation-equivariant function, and that any univariate query-document scoring function can be extended to capture cross-document interactions using this proposed self-attention mechanism. We choose the attention method used in Transformer~\cite{vaswani2017attention} in our paper and combine the output of self-attention layers with a feed forward network in a wide and deep architecture. We conducted our experiments on three datasets and the results show that our proposed methods can capture document interactions effectively in a statistically significant manner, and can efficiently scale to large document sets.

\balance
\bibliographystyle{plainnat}
\bibliography{main} 
\end{document}